# *Field Electron Emission Theory (October 2016),v2*

**Richard G. Forbes**

Advanced Technology Institute & Department of Electrical and Electronic Engineering,
Faculty of Engineering & Physical Sciences, University of Surrey,
Guildford, Surrey GU2 7XH, UK
Permanent e-mail alias:  r.forbes@trinity.cantab.net

**Explanatory note added January 2018**

This paper is based on the material in two tutorial lectures originally delivered at the 2016 Young Researchers School in Vacuum Micro- and Nanoelectronics, held at Saint-Petersburg Electrotechnical University ("LETI"), on 5th-6th October 2016.

The original version of this paper was published electronically as part of the School Proceedings, via the IEEE Explore Digital Library (see Ref. [24]).

This version (v2) incorporates some improvements and updates made in May 2017 (see note at end), and some further small improvements (mainly as regards terminology) made in January 2018.

Updated versions of the original tutorial lectures have been delivered at an International Conference in Jordan in April 2017, and at the 2017 Young Researchers School in Vacuum Micro- and Nanoelectronics, again held at Saint-Petersburg Electrotechnical University ("LETI"), on 5th-6th October 2017. Links to the related powerpoint presentations are given as Refs [26] and [27].

# Field Electron Emission Theory (October 2016)

Richard G. Forbes

Advanced Technology Institute & Department of Electrical and Electronic Engineering,
Faculty of Engineering & Physical Sciences, University of Surrey,
Guildford, Surrey GU2 7XH, UK
Permanent e-mail alias: r.forbes@trinity.cantab.net

*Abstract*—This conference paper provides an overview of the material presented in two field electron emission tutorial lectures given at the 2016 Young Researchers' School in Vacuum Micro- and NanoElectronics, held in Saint-Petersburg in October 2016. This paper aims to indicate the scope and structure of the tutorials, and also where some of the related published material can be found.

*Keywords—field electron emission; Fowler-Nordheim theory; principal Schottky-Nordheim barrier function; Gauss variable.*

INTRODUCTION

*A. General introduction*

This paper provides an overview of the material presented in two field electron emission tutorials [1,2] given at the 2016 Young Researchers' School in Vacuum Micro- and Nano-Electronics, held in Saint-Petersburg in October 2016.

Obviously, much of the scientific content of the tutorials has already been published. This paper aims to indicate the scope and structure of the tutorials, and also where some of the related published material can be found.

To enhance the clarity of non-oral presentations, small amounts of additional material have been included, and the content has been slightly re-ordered.

*B. Introductory issues*

These tutorials are principally about mainstream field electron emission (FE) theory, particularly theory associated with the derivation and application of Fowler-Nordheim-type (FN-type) equations. This topic is sometimes called *cold field electron emission (CFE)*, sometimes *Fowler-Nordheim field electron emission (FNFE)*. I now prefer the second name. Theory represents understanding, as well as mathematics. Understanding FE theory is of enduring importance.

Strictly, mainstream FE theory applies to *metal* emitters that are "not too sharp" (apex radius > 10-20 nm). But it is often applied to other situations, as a first approximation. Presumably, when this works, it does so because barrier effects are often the most important effects in FNFE, and surface barrier behavior can be fairly similar for metallic and non-metallic materials.

II. BASIC CONCEPTS AND CONVENTIONS

*A. Equations systems and unit systems*

The SI unit system [3] is based on a system of quantities and equations (involving the electric constant $\varepsilon_0$) that, since 2009, has been called [3] the *International System of Quantities (ISQ)*. Modern FE theory uses ISQ equations, but a system of customary FE units [4] based on the eV and the V/nm.

*B. The electron emission convention*

Conventional FE is induced by an *electrostatic* field that is negative in value. The *electron emission convention* (used here) is to treat all fields, currents and current densities as positive, even though they would be negative in classical electrostatics.

The conventional symbol for classical electrostatic field is ***E***. To avoid confusion, many FE theoreticians use the symbol ***F*** (or *F*) to denote the negative or magnitude of classical electrostatic field. This is done here, and leaves the symbol *E* free to denote electron energy. However, many experimental papers use the symbol *E* to represent a positive quantity that is the absolute magnitude of classical electrostatic field ***E***.

*C. The Sommerfeld model*

In the Sommerfeld emitter model, the total electron energy $E_t$ can be split into components parallel ($E_p$) and normal ($E_n$) to the emitter surface, with $E_t = E_n + E_p$. I call the direction normal to the surface the *forwards direction*, and $E_n$ the *forwards energy*. In emission contexts, the vertical axis of diagrams illustrating the Sommerfeld model represents forwards energy.

Note the distinction between (a) the *Fermi level (FL)*, which (in the Sommerfeld model) is the total-energy level at which the Fermi-Dirac occupation probability has value 0.5, and (b) the *Fermi energy*, which is the total-energy difference between the Fermi level and the base of the conduction band.

Total and forwards energies can be measured relative to any convenient reference level, but the same reference level must be used for both. It is often convenient to use the Fermi level as the reference, in which case the symbol *E* is replaced by $\varepsilon$.

When a constant external electrostatic field of magnitude *F* is applied to the Sommerfeld model, this creates the *exactly triangular (ET) barrier*, in which the variation (for *z*>0) of the

*electrostatic component* $U^{ES}$ (measured relative to the Fermi level) *of the electron potential energy* is given by

$$U^{ES} = \phi - eFz, \quad (1)$$

where $\phi$ is the local work function, $e$ the elementary positive charge, and $z$ distance measured from the well edge.

### D. Work-function theory and patch fields

Values of local work-function $\phi$ (and also the local Sommerfeld-well depth) are due to two components, related to (a) bulk chemical effects and (b) "chemically-induced" surface-electric-dipole effects. The former causes $\phi$-differences between materials, the latter $\phi$-differences between different crystal faces of a given material [5]. The surface-dipole effects are due to a balance between electron *spreading* into the vacuum (considered the same for all faces) and electron *smoothing* sideways into the gaps between surface atoms (which depends on surface crystallographic structure).

When different parts ("patches") of an emitter surface have different work functions, a system of electrostatic *patch fields* [6,7] exists outside the surface, strongest near the patch edges.

### E. Elements of charged-surface theory

With a real, flat, planar, atomically-structured emitter surface, one wants $U^{ES}(z)$ to have form (1) at large distances from the surface. The issue is "how to determine the position of the plane in which $z$ has the value zero?" This plane is termed the emitter's *electrical surface* [8].

At a real charged surface, the surface atoms are *polarized*, i.e., they have a field-induced electric dipole moment. This is a *universal* property of solid materials, including metals. For a planar surface, it is readily shown [8] that, if the atoms were not polarized, then the electrical surface would lie in the plane of the surface-atom nuclei. The effect of the surface-atom polarization is to repel the electrical surface outwards, towards the vacuum, by a distance $d_{rep}$ called the *repulsion distance*.

This effect can be modeled using classical electrostatics and a *classical array model* in which point charges and polarizable point dipoles are placed at the positions of the surface-atom nuclei, with polarizability values taken from the literature. It is found that $d_{rep}$ is approximately equal to the relevant atomic radius, as assessed by half the nearest-neighbor distance.

In those cases where comparisons can be made, these classical results are very close (within 20 pm or better) to advanced quantum-mechanical calculations [8,9], and are also compatible with appearance-energy experiments [10].

The Sommerfeld model is fitted to emission at a real planar surface by putting the well edge in the electrical surface.

### F. Image potential energy and the Schottky-Nordheim barrier

In the absence of any external electrostatic field, an electron just outside the emitter has an *exchange-and-correlation (XC)* interaction with the emitter surface. In FE, this is usually modeled as given by Schottky's [11] classical planar image potential energy $U^{im} = -e^2/16\pi\varepsilon_0 z$. Adding this to $U^{ES}$ yields the *total electron potential energy* $U^{tot}$ (relative to the FL) given by

$$U^{tot} \approx U^{ES} + U^{im} = \phi - eFz - e^2/16\pi\varepsilon_0 z. \quad (2)$$

This PE variation is sometimes called the *Schottky-Nordheim (SN) PE barrier*.

As compared with the ET barrier, the height of the SN barrier is reduced by an amount $\Delta$ given by

$$\Delta = c_S F^{1/2} \equiv (e^3/4\pi\varepsilon_0)^{1/2} F^{1/2}, \quad (3)$$

where $c_S [\equiv (e^3/4\pi\varepsilon_0)^{1/2}]$ is the *Schottky constant* [4].

For a barrier of zero-field height $H$, the *reference field* $F_{R,H}$ necessary to reduce the barrier to zero is $F_{R,H} = c_S^{-2} H^2$. For this barrier of zero-field height $H$, the *scaled barrier field* $f_H$ is defined by

$$f_H \equiv F/F_{R,H} = c_S^2 H^{-2} F. \quad (4a)$$

For a barrier of zero-field height equal to the local work function $\phi$, the reference field is denoted simply by $F_R$, the related scaled barrier field is denoted simply by $f$, and eq. (4a) becomes

$$f \equiv F/F_R = c_S^2 \phi^{-2} F. \quad (4b)$$

This dimensionless parameter $f$ plays an important role in modern FE theory.

### G. Overview of emission concepts

In FNFE (CFE), almost all electrons escape by *wave-mechanical tunneling* through a field-lowered energy barrier, from states close in energy to the Fermi level. Contrary to what is often stated, tunneling is **not** a mysterious counter-intuitive effect. It is simply a *classical* mathematical property of wave-equations. For electrons this is the Schrödinger equation. But tunneling can also occur with light, with sound, and with waves on strings, all of which obey wave-equations.

Consider an electron, in a wave-mechanical state $\mathbf{k}$, approaching the tunneling barrier at the emitter surface. The probability that the electron will escape (rather than be reflected) is termed the *transmission probability* (or tunneling probability) and is denoted by $D_k$.

Associated with each travelling-wave state $\mathbf{k}$ there is a contribution $z_k$ to the *incident* current density $Z$ approaching the surface from the inside, and a contribution $z_k D_k$ to the *emission* current density (ECD). The total *local ECD* $J$ is obtained by summing over the incident states. Formally:

$$J = \Sigma_k (z_k D_k). \quad (5)$$

In practice, this summation is usually best done by means of a double integration with respect to energy components. Thus, as preliminaries we need: (a) a theory of electron-state energies, and of the distribution of electrons as between different energy states; and (b) a theory of transmission probability.

### H. Elements of free-electron theory

Inside the emitter, the contribution $d^2 Z$ (to incident current density) arising from states in the energy range $d\varepsilon_n d\varepsilon_p$ is

$$d^2 Z = z_{sup} \, d\varepsilon_n d\varepsilon_p, \quad (6)$$

where $z_{sup}$ is the relevant supply density. (The term *supply*

*density* is defined to mean the electron current density approaching the emitter surface *per unit area of energy-space*.)

When all energy-states in energy range $d\varepsilon_n d\varepsilon_p$ are taken as fully occupied, the supply density is constant in energy-space and equal to the *Sommerfeld supply density* $z_S$ given by [12]

$$z_S = 4\pi e m_e/h_P^3 \approx 1.618311 \times 10^{14} \text{ A m}^{-2}\text{eV}^{-2}, \quad (7)$$

where $m_e$ is the electron mass in free space and $h_P$ is Planck's constant. This is a fundamental statistical-mechanical result that is exactly equivalent to assuming that "the density of electron states is constant in phase-space", but is a better starting point for emission theory.

It is normally assumed that the emitter electron states can be treated as in local thermodynamic equilibrium, with occupation probability given by the *Fermi-Dirac distribution function*.

$$f_{FD} = 1/[1+\exp\{\varepsilon_t/k_B T\}] = 1/[1+\exp\{(\varepsilon_n+\varepsilon_p)/k_B T\}], \quad (8)$$

where $k_B$ is the *Boltzmann constant* and $T$ is *thermodynamic temperature*.

Hence the local ECD becomes given by the integral

$$J = z_S \iint f_{FD}(\varepsilon_t) D(\varepsilon_n) d\varepsilon_n d\varepsilon_p. \quad (9)$$

Since $D$ is a function only of $\varepsilon_n$, we want to convert this to an integral of the form

$$J = \int j_n(\varepsilon_n) d\varepsilon_n = \int N(\varepsilon_n) D(\varepsilon_n) d\varepsilon_n, \quad (10)$$

where $j_n(\varepsilon_n)$, [usually called the *(emitted) normal energy distribution*] is the product of $D(\varepsilon_n)$ and the quantity $N(\varepsilon_n)$, which is accordingly given by

$$N(\varepsilon_n) \equiv z_S \int_0^\infty [1+\exp\{(\varepsilon_n+\varepsilon_p)/k_B T\}]^{-1} d\varepsilon_p, \quad (11)$$

$$N(\varepsilon_n) = z_S k_B T \ln[1+\exp\{-\varepsilon_n/k_B T\}]. \quad (12)$$

In the past, $N(\varepsilon_n)$ has sometimes been called the *current supply function* [or the theory has employed an alternative quantity equal to $N(\varepsilon_n)/e$. However, a better name for $N(\varepsilon_n)$ as used here is the *incident normal energy distribution* (or "incident NED"). In the limit of high temperature, eq. (12) becomes

$$N(\varepsilon_n) \approx z_S k_B T \exp\{-\varepsilon_n/k_B T\}. \quad (13)$$

Strictly, integral (10) over $\varepsilon_n$ runs from the bottom of the conduction-band to $+\infty$. But usually there is no significant emission near the band base, so formally the lower limit can be extended to $-\infty$. Thus, integral (10) can be replaced by

$$J_T = \int_{-\infty}^{+\infty} N(\varepsilon_n) D(\varepsilon_n) d\varepsilon_n, \quad (14)$$

where $J$ has been replaced by $J_T$, to show that a finite-temperature expression is being derived.

I. *Transmission probability theory*

Exact solution of the Schrödinger equation is not possible for most rounded barriers, so the following approximate "semi-classical" approach is used. The Schrödinger equation is separated in Cartesian coordinates, and the component relating to motion normal to the emitter surface is written in the form

$$[\hat{K}_z + \{U^{tot}-\varepsilon_n\}]\Psi_z = [\hat{K}_z + M(z)]\Psi_z = 0, \quad (15)$$

where $\hat{K}_z$ is the relevant kinetic-energy operator, and the *electron motive energy* $M(z)$ is defined by $M(z) \equiv \{U^{tot}(z)-\varepsilon_n\}$. The motive energy for an SN barrier corresponding to a state with forwards energy $\varepsilon_n$ is thus given by

$$M^{SN}(z) = (\phi-\varepsilon_n)-eFz-e^2/16\pi\varepsilon_0 z \equiv H-eFz-e^2/16\pi\varepsilon_0 z, \quad (16)$$

where the *zero-field barrier height* $H \equiv (\phi-\varepsilon_n)$. In general, a parameter $G$ called the *barrier strength* can then be defined by

$$G \equiv g_e \int M^{1/2}(z) dz, \quad (17)$$

where $g_e$ is the *JWKB constant for an electron* [4], and the integral is taken "across the barrier" [i.e., between the relevant zeros of $M(z)$]. Strong barriers are difficult to tunnel through.

For the exactly triangular (ET) barrier, the barrier strength $G^{ET}$ is $bH^{3/2}/F$, where $b$ is the *second FN constant* given by [4]

$$b = 2g_e/3e = (4/3)(2m_e)^{1/2}/e\hbar, \quad (18)$$

where $\hbar$ is Planck's constant divided by $2\pi$.

For any other barrier, [i.e. the "general barrier" (GB), with motive energy $M^{GB}(z)$], we can define a *barrier form correction factor* $v^{GB}$ ("nu$^{GB}$") via

$$G^{GB} = g_e \int [M^{GB}(z)]^{1/2} dz \equiv v^{GB} G^{ET} = v^{GB} bH^{3/2}/F. \quad (19)$$

In semi-classical quantum theory, several different approximate formulae exist for relating the transmission probability $D$ for a given barrier to the related barrier strength $G$. In practice, advanced FE treatments normally use the *Kemble formalism/approximation*

$$D \approx 1/\{1+\exp G\}. \quad (20)$$

When the barrier is sufficient strong, i.e. $G>(3$ to $5)$, eq. (20) reduces adequately to the *simple-JWKB formalism/approximation*

$$D \approx \exp[-G]. \quad (21)$$

J. *Decay width*

The *decay width d* is a (positive) measure of how quickly $D$ falls off with increase in barrier height, and is defined via

$$d^{-1} = -(\partial \ln D/\partial H)_F = (\partial \ln D/\partial \varepsilon_n)_F. \quad (22)$$

When these derivatives are evaluated for $H=\phi$ ($\varepsilon_n=0$), the resulting parameter is denoted by $d_F$ and called the *decay width at the Fermi level*.

When the simple-JWKB approximation is used, $d^{-1}$ becomes given by

$$d^{-1} \approx (\partial G/\partial H)_F = -(\partial G/\partial \varepsilon_n)_F, \quad (23)$$

and more detailed (approximate) formulae can be found for $d_F$.

(1) For the ET barrier:

$$d_F^{ET} \approx (2/3b) \phi^{-1/2} F. \quad (24)$$

Values are typically around 0.2 eV to 0.3 eV.

(2) For the general barrier, $d_F^{GB}$ is (for historical reasons) related to $d_F^{ET}$ via a *decay-rate correction factor* $\tau_F^{GB}$, with

$$d_F^{GB} \approx (\tau_F^{GB})^{-1} d_F^{ET}, \quad (25a)$$

$$\tau^{GB} \equiv v^{GB} + (2/3) H (\partial v^{GB}/\partial H)_F. \quad (25b)$$

(3) For the SN barrier, we can make use of eq. (4a), subject to the condition that $F$ is held constant, to convert eq. (25b) to a form involving $f_H$, namely

$$\tau^{SN} \equiv v^{SN} - (4/3) f_H \, dv^{SN}/df_H. \quad (25c)$$

For a barrier of zero-field height $\phi$, the notation is simplified and this becomes

$$\tau_F^{SN} \equiv v_F^{SN} - (4/3) f \, dv^{SN}/df. \quad (25d)$$

This result enables us to see that $\tau_F^{SN}$ is given by an appropriate particular value $t_F$ [=t($f$)] of a special mathematical function t($x$) discussed below and defined by

$$t(x) \equiv v(x) - (4/3)x \, dv/dx, \quad (26)$$

where $x$ is the *Gauss variable* [see Section IV below].

*K. Emission current density regimes*

In integral (14), the product $N(\varepsilon_n)D(\varepsilon_n)$ appears in the integrand. Results above show that both $N(\varepsilon_n)$ and $D(\varepsilon_n)$ have "full" and "approximate" forms [equations (12) and (13), and (21) and (22), respectively]. Integral (14) cannot be exactly evaluated analytically if both "full" forms are used, although (obviously) it can be evaluated numerically.

To obtain a simple analytical result, at least one of $N(\varepsilon_n)$ and $D(\varepsilon_n)$ must be approximated. This gives rise to the idea of *emission current density (ECD) regimes* (also called "emission regimes") where a simple analytical ECD formula can be obtained. Of special interest here is the *FNFE (or CFE) regime*, defined by using the "approximate form" [eq. (21)] for $D(\varepsilon_n)$ and the "full form" [eq. (12)] for $N(\varepsilon_n)$. As first shown by Murphy and Good [13], the resulting mathematics leads to finite-temperature Fowler-Nordheim-type equations.

Using eq. (20) for $D(\varepsilon_n)$ and eq. (14) for $N(\varepsilon_n)$ leads to what the author calls the *barrier-top electron emission* (BTE) regime (also called the "extended Schottky regime" [14]).

Note that neither regime corresponds to the physical phenomenon called *thermal electron emission (TE)* (also called "thermionic emission"). In fact, "wave-mechanical" and "classical" TE regimes can be defined by taking eq. (14) for $N(\varepsilon_n)$ and other expressions for $D(\varepsilon_n)$. There are also special high-field regimes, for example "explosive emission" and "liquid-metal electron source" regimes.

There is also a "general thermal-field formula", proposed by Jensen and Cahay [15,16], that provides good approximate results across wide ranges of field and temperature.

## III. DERIVATION OF FINITE TEMP. FN-TYPE EQUATIONS

*A. The temperature correction factor $\lambda_T$*

To carry out this derivation, we use eqs (14), (12) and (21), but first must expand $\ln\{D(\varepsilon_n)\}$ about the Fermi level [$\varepsilon_n=0$]. In the simple-JWKB formalism/approximation, we have

$$\ln\{D(\varepsilon_n)\} = -G(\varepsilon_n)|_0 - (\partial G/\partial \varepsilon_n)|_0 \varepsilon_n - \tfrac{1}{2}(\partial^2 G/\partial \varepsilon_n^2)|_0 \varepsilon_n^2 - \ldots \quad (27)$$

Neglecting terms of second order and higher, then using the eq. (24) definition of decay width, and using subscript "F" to label values "taken at the Fermi level", yields

$$\ln\{D(\varepsilon_n)\} \approx -G_F + \varepsilon_n/d_F, \quad (28a)$$

$$D(\varepsilon_n) \approx D_F \exp[\varepsilon_n/d_F]. \quad (28b)$$

Hence, integral (14) becomes

$$J_T = z_S D_F \int_{-\infty}^{+\infty} k_B T \ln[1+\exp\{-\varepsilon_n/k_B T\}] \cdot \exp[\varepsilon_n/d_F] \, d\varepsilon_n. \quad (29)$$

At this point it is useful to follow the approach of Swanson and Bell [16], and introduce a parameter $p$ defined by

$$p \equiv k_B T/d_F. \quad (30)$$

Provided $p<1$ (which is certainly true physically under most real circumstances), eq. (29) can be integrated by parts, to yield

$$J_T = z_S d_F D_F \int_{-\infty}^{+\infty} [\{\exp(\varepsilon_n/d_F)\} / \{1+\exp(\varepsilon_n/k_B T)\}] \, d\varepsilon_n. \quad (31)$$

This integral is a standard form found in tables, and yields

$$J_T = \lambda_T z_S d_F^2 D_F = \{(\pi p)/\sin(\pi p)\} z_S d_F^2 D_F, \quad (32)$$

where $\lambda_T [\equiv (\pi p)/\sin(\pi p)]$ is a *temperature correction factor* defined by eq. (32). In the zero-temperature limit: $\lambda_T \to 1$, $J_T$ becomes equal to the zero-temperature expression $J_0$, and we can write

$$J_T = \lambda_T J_0 = \lambda_T (z_S d_F^2 D_F). \quad (33)$$

Although originally formulated by Murphy and Good [13] in the context of the SN barrier, the argument and result as presented here in fact apply to a barrier of any well-behaved form, as pointed out in Ref. [17].

*B. Abstract form for a FN-type equation*

The form $J_0 = z_S d_F^2 D_F$ is, in fact, a useful abstract form for a zero-temperature Fowler-Nordheim-type equation. The quantity $d_F^2$ has a useful interpretation as the "effective area in energy-space from which emitted electrons are drawn", and the quantity $z_S d_F^2$ has a useful interpretation as the "effective supply" ["effective incident current density ($Z_F$)"] of electrons onto a barrier of zero-field height $\phi$.

By using eqs (18), (24) and (25a), it can be shown that

$$z_S d_F^2 = (\tau_F^{GB})^{-2} a\phi^{-1} F^2, \quad (34)$$

$$J_0 = (\tau_F^{GB})^{-2} a\phi^{-1} F^2 \exp[-v_F^{GB} b\phi^{3/2}/F], \quad (35)$$

where $a$ is the *first FN constant*, given by [4]:

$$a = e^3/8\pi h_P, \quad (36)$$

where $h_P$ is Planck's constant.

Equation (35) is a slightly generalized form of the Murphy-Good result. As already indicated, for the SN barrier the correction factors $v_F^{SN}$ and $\tau_F^{SN}$ are given by parameters $v_F$ and $t_F$ that are appropriate particular values of special mathematical functions v($x$) and t($x$). The next section presents relevant background theory.

## IV. THE SPECIAL MATHEMATICAL FUNCTION v($x$)

### A. Mathematical background

The special mathematical function v (usually $\vartheta$ in Russian texts), which I currently call the either the *principal SN barrier function* or the *field emission* v-*function*, has many equivalent mathematical definitions, but the most fundamental is as a special solution [18] of the Gauss Hypergeometric Differential Equation (HDE):

$$x(1-x)d^2W/dx^2 + [c_G - (a_G + b_G + 1)x]dW/dx - a_G b_G W = 0 , \quad (37)$$

where $a_G$, $b_G$ and $c_G$ are constants. We can call $x$ the *Gauss variable*. Taking $a_G = -3/4$, $b_G = -1/4$, $c_G = 0$, reduces eq. (37) to the special equation identified by Forbes and Deane (see [18]), namely

$$x(1-x)d^2W/dx^2 = (3/16) W . \quad (38)$$

This is a special mathematical equation, like Airy's and Bessel's equations, but is much more obscure. v($x$) is a particular solution of eq. (38) satisfying the boundary conditions [18]:

$$v(0) = 1; \quad \lim_{x \to 0} \{dv/dx - (3/16)\ln x\} = -(9/8)\ln 2 . \quad (39)$$

Exact analytical and series expressions for v($x$) are known [18], and a "good simple approximation" has been found [19], namely

$$v(x) \approx 1 - x + (1/6)x\ln x . \quad (40)$$

This has accuracy of 0.33% or better over the range $0 \leq x \leq 1$ [4]. The function v($x$) as defined by eqs (38) and (39) exists in the range $0 \leq x \leq \infty$, but the accuracy of eq. (40) deteriorates increasing rapidly for $x > 1$, and approximation (40) should not be used outside the range $0 \leq x \leq 4$.

v($x$) also has integral definitions. One useful definition [13, 20] is

$$v(x) = (3 \times 2^{-3/2}) \int_{b'}^{a'} (a'^2 - \eta^2)^{1/2} (\eta^2 - b'^2)^{1/2} d\eta , \quad (41)$$

where: $a' = \{1 + (1-x)^{1/2}\}^{1/2}$; $b' = \{1 - (1-x)^{1/2}\}^{1/2}$, and all square roots are positive. Primes have been added to the symbols "$a$" and "$b$", in order to distinguish them from the FN constants.

Form (41) is a standard form that can be expressed (in many ways) in terms of complete elliptic integrals of the first [$K(m)$] and second [$E(m)$] kinds. Here, $m$ is the *elliptic parameter* (i.e., $m = k^2$, where $k$ is the elliptic modulus). If $m(x)$ is taken as

$$m(x) = (1 - x^{1/2})/(1 + x^{1/2}) , \quad (42)$$

then v($x$) has the elliptic-integral expression/definition [13,20]

$$v(x) = (1 + x^{1/2})^{1/2} [E(m(x)) - x^{1/2} K(m(x))] . \quad (43)$$

This form is useful for entry into computer-algebra packages.

Given v($x$), a derived special mathematical function t($x$) can be defined via eq. (26).

### B. Application to modelling barrier transmission

The aim of this section is to show why the function v($x$) is relevant to evaluating transmission coefficients when using the simple-JWKB formalism/approximation. Although first used in the context of the SN barrier, the function v($x$) is actually relevant to modeling a more general type of barrier, with motive energy

$$M(z) = H - eFz - \beta/z , \quad (44)$$

where $\beta$ is a constant. For example, a barrier of this form occurs field ionization theory. Form (44) has no established name. I call it a *basic Laurent-form barrier*. It is convenient to first give this more general theory.

Simple semi-classical ("JWKB-type") tunneling theory involves an integral of $M^{1/2}(z)$ between adjacent zeros. The zeros of eq. (44) occur at

$$z_+, z_- = (H/2eF)\{1 \pm (1-\mu)^{1/2}\} \equiv (H/2eF)\{1 \pm \alpha\} \quad (45)$$

where the *modeling parameter* $\mu$ is given by

$$\mu \equiv 4\beta eF/H^2 = F/F_R , \quad (46)$$

and

$$\alpha \equiv (1-\mu)^{1/2} . \quad (47)$$

The reference field $F_R$ [$= H^2/4e\beta$] is the field that makes the barrier peak height equal to zero. $\alpha$ is the same parameter as used in Ref. [20], but has a different formal definition.

By arguments that are exactly analogous to those in Ref. [20], it may be shown [from eq. (19) there] that the simple-JWKB transmission probability for barrier (44) is given by the linked equations

$$D = \exp[-\{(4/3)(2m_e)^{1/2}/e\hbar\} \{H^{3/2}/F\} \cdot I_{\text{new}}]$$
$$= \exp[-\{bH^{3/2}F\} \cdot I_{\text{new}}] , \quad (48)$$

$$I_{\text{new}} = (3 \times 2^{-3/2}) \int_{b'}^{a'} (a'^2 - \eta^2)^{1/2} (\eta^2 - b'^2)^{1/2} d\eta , \quad (49)$$

with

$$a' = \{1 + \alpha\}^{1/2} = \{1 + (1-\mu)^{1/2}\}^{1/2}, \quad (50)$$

$$b' = \{1 - \alpha\}^{1/2} = \{1 - (1-\mu)^{1/2}\}^{1/2} . \quad (51)$$

Comparison of eqs (49) to (51) with definition (41) for v($x$) given earlier shows that $I_{\text{new}}$ can be obtained by setting $x = \mu$ in v($x$), and consequently that

$$D = \exp[-v(\mu) bH^{3/2}/F] . \quad (52)$$

In the case of the SN barrier of zero-field height $\phi$, the modeling parameter $\mu$ is given by the scaled barrier field $f$ [for a SN barrier of zero-field height $\phi$], and the transmission coefficient at the Fermi level, $D_F^{SN}$, is given by

$$D_F^{SN} = \exp[-v(f) b\phi^{3/2}/F] . \quad (53)$$

It follows that the corresponding zero-temperature ECD equation is:

$$J_0 = \{t(f)\}^{-2} a\phi^{-1} F^2 \exp[-v(f) b\phi^{3/2}/F] . \quad (54)$$

Equation (54) is the *Murphy-Good zero-temperature FN-type equation*. Note that, for clarity, the symbols v($f$) and t($f$) are often replaced in the literature by the equivalent symbols $v_F$ and $t_F$ (see below).

Two general points emerge from this discussion.
(1) The mathematics of v($x$) is valid pure mathematics in its own right, and applicable to all basic Laurent-form barriers. It

would be useful to make a firmer distinction between the pure-mathematics aspects of v(x) and its applications in modeling.

(2) It is also useful to distinguish: (i) the pure-mathematical (Gauss variable) $x$; (ii) the general Laurent-form-barrier-modeling variable $\mu$; and (iii) the particular modeling variables (in particular $f$) used in modeling the SN barrier.

*C. Why change from using the Nordheim parmeter y ?*

Older FE literature uses the *Nordheim parameter y* $[=+\sqrt{x}]$ as the argument of v. There are good mathematical and pragmatic reasons for now using $x$, $\mu$, $f_H$ and $f$, rather than $y$.

(1) The natural pure-mathematical variable to use is the Gauss variable $x$: the modeling equivalents of $x$ are $\mu$, $f_H$ and $f$ (rather than $y$).

(2) No terms in $x^{1/2}$ appear in the exact series expansion for v(x).

(3) The defining equation for v is simpler when written in terms of $x$, rather than in terms of $y$.

(4) The variable $f$ is proportional to the barrier field $F$, whereas $y$ is proportional to $\sqrt{F}$. This makes $f$ easier to use than $y$, particularly in the context of an orthodoxy test [21].

(5) The concept of "scaled barrier field" is probably easier to understand and use than the Nordheim parameter (which is actually "scaled reduction in SN-barrier height").

(6) The symbol $f$ has a unique definition, whereas, historically, the symbol $y$ has had several related (slightly different) meanings.

(7) The concept of "scaled barrier field" is, in principle, more general than the Nordheim-parameter concept, and can be extended (when suitably modified in detail) to apply both to general modeling barriers and to real physical barriers.

(8) The parameter $f$ seems likely to prove more generally useful than $y$.

It would make for a simpler and clearer system if use of the variable $y$ were phased out, and relevant formulas and tables using $y$ were replaced by formulas and tables using $x$ or $f$.

In the present mixed system, a danger for non-experts is confusion over the meanings of "v(f)" and "v(y)", as this is not a change in symbol but a change in variable. To avoid this confusion, the argument-free symbols $v_F$ and $t_F$ are sometimes used in the Murphy-Good FN-type equation.

## V. BRIEF HISTORY OF FIELD ELECTRON EMISSION

This material [2] has been presented on several previous occasions, and is not discussed in detail here. For mainstream theory, the main historical phases seem to have been:

1745-1923: Phenomenological phase
1923-1928: The search for linearity in experimental data
1928-1950s: The Fowler-Nordheim phase
1950s-1990s: The Murphy-Good phase
1990s-present: Reconstruction phase.

I hope that we can soon move on to a "more scientific" phase.

A review of experimental work, given in the second tutorial [2], enables us to develop views as what theory is, in fact, needed to support experimental FE activity.

## VI. APPLYING FOWLER-NORDHEIM-TYPE THEORY

These sections discuss a few of the conceptual and related issues that arise when attempting to apply basic FN-type theory to real experimental situations. It is not intended as a complete discussion.

*A. The concept of a Fowler-Nordheim-type equation*

A *Fowler-Nordheim-type equation* is any FNFE equation with the mathematical form:

$$Y = C_{YX}X^2 \exp[-B_X/X] , \qquad (55)$$

where: $X$ is any FE independent variable (usually a voltage or a field); $Y$ is any FE dependent variable (usually a current or current density); $B_X$ is a variable parameter related to the choices of $X$ and barrier form; and $C_{YX}$ is a variable parameter related to the choices of $Y$ and $B_X$.

The *core theoretical forms* of FN-type equations (those derived most directly from basic theory) give the local ECD $J_L$ in terms of local work function $\phi$ and the local barrier field $F_L$.

The parameters $\phi$, $F_L$ and $J_L$ all vary with position on the emitter surface. To derive an formula for emission current, it is necessary to consider a *characteristic point* "C" on the emitter surface, and to denote the *characteristic* barrier-field and ECD values there by $F_C$ and $J_C$, respectively. In modeling, it is usual to take the characteristic point at *the emitter apex*, since the local barrier field is highest there (if patch-field effects are not present or are disregarded) Also, in modeling, it is often convenient to take the core form of a FN-type equation to give $J_C$ in terms of $\phi$ and $F_C$. The remarks that follow here refer to the core forms of equation, as defined in this way.

(1) The simplest type of core FN-type equation is the so-called *elementary FN-type equation*:

$$J_C^{el} = a\phi^{-1}F_C^2 \exp[-b\phi^{3/2}/F_C] . \qquad (56)$$

This is based on assuming an exactly triangular (ET) tunneling barrier, and is a simplified form of the original (1928) FN-type equation.

(2) As indicated previously, the ET barrier is considered not physically satisfactory because it disregards XC ("image") effects, and is not adequately valid for highly curved surfaces. Inclusion of a *barrier form correction factor* $v_F^{GB}$ leads to the equation form

$$J_{kC}^{GB} = a\phi^{-1}F_C^2 \exp[-v_F^{GB}b\phi^{3/2}/F_C]. \qquad (57)$$

I call this the *kernel current density* for the general barrier (GB). For given values of $v_F^{GB}$, $\phi$ and $F_C$, values for the kernel current density $J_{kC}^{GB}$ can be evaluated *precisely*.

(3) To allow for other corrections, which may have multiple causes [1], it is necessary to introduce a *pre-exponential correction factor* $\lambda_C^{GB}$, so that the equation becomes

$$J_C^{GB} = \lambda_C^{GB}J_{kC}^{GB} = \lambda_C^{GB}a\phi^{-1}F_C^2 \exp[-v_F^{GB}b\phi^{3/2}/F_C] . \quad (58)$$

This is the *most general form* of FN-type equation (though it is not the most general form of equation that could be devised to describe FNFE).

A major problem for FE science is that the value of $\lambda_C$ is not reliably known for any physically realistic barrier model.

Thus, for the SN barrier it is guessed [22] that $\lambda_C^{SN}$ lies somewhere in the range $0.005 < \lambda_C^{SN} < 11$.

### B. Equation complexity levels

Historically, many different assumptions have been used to obtain many different expressions for $\nu_F^{GB}$ and $\lambda_C^{GB}$. The choices of assumptions and expressions determine the *complexity level* of the resulting approximate equation.

TABLE I. Complexity levels of Fowler-Nordheim-type equations that apply to a flat, smooth, planar emitter surface.

| Name | Date | $\lambda_C^{GB} \rightarrow$ | Barrier form | $\nu_F^{GB} \rightarrow$ | Note |
|---|---|---|---|---|---|
| Elementary | ? | 1 | ET | 1 | a |
| Original | 1928 | $P_F^{FN}$ | ET | 1 | b |
| Fowler-1936 | 1936 | 4 | ET | 1 | |
| Extended elementary | 2015 | $\lambda_C^{ET}$ | ET | 1 | |
| Dyke-Dolan | 1956 | 1 | SN | $\nu_F$ | |
| Murphy-Good zero-temperature | 1956 | $t_F^{-2}$ | SN | $\nu_F$ | |
| Murphy-Good finite-temperature | 1956 | $\lambda_T t_F^{-2}$ | SN | $\nu_F$ | |
| Orthodox | 2013 | $\lambda_C^{SN0}$ | SN | $\nu_F$ | c |
| New-standard | 2015 | $\lambda_C^{SN}$ | SN | $\nu_F$ | |
| "Barrier-effects-only" | 2013 | $\lambda_C^{GB0}$ | GB | $\nu_F^{GB}$ | c |
| General | 1999 | $\lambda_C^{GB}$ | GB | $\nu_F^{GB}$ | |

<sup>a</sup>Many earlier imprecise versions exist, but the first clear statement seems to be in 1999 [23].
<sup>b</sup>For details concerning the Fowler-Nordheim tunnelling pre-factor $P_F^{FN}$, see [4].
<sup>d</sup>The superscript "0" indicates that the factor is to be treated mathematically as constant.

Even for emitters assumed to have smooth classical *planar* surfaces, many different complexity levels exist, as shown in Table I above. The view of the author is that we should consider the so-called "new-standard" FN-type equation as the equation best currently suited to discuss experimental results.

### C. Auxiliary equations

In order to relate other variables to the core variables $F_C$ and $J_C$ (or, better, $J_{kC}$), it is necessary to introduce *auxiliary equations*. These have the general forms

$$F_C = c_X X, \quad (59)$$
$$Y = c_Y J_{kC}, \quad (60)$$

where $c_X$ and $c_Y$ are *auxiliary parameters*.

A detailed discussion is given in Ref. [22]. The most important cases are those where $X$ is the emission voltage $V_e$, and $Y$ is the emission current $i_e$. With eq. (59) there are alternative options, and confusion also exists in the literature over nomenclature (see [21,22]). The author's preferred option is now to write

$$F_C = V_e/\zeta_C, \quad (61)$$

where $\zeta_C$ is the characteristic *local conversion length* (LCL).

Another significant case is planar-parallel-plate geometry, where the *true macroscopic field* $F_M$ can be written in the form

$$F_M = V_e/\zeta_M, \quad (62)$$

and (if the emission situation is "orthodox" – see below) the *macroscopic conversion length* $\zeta_M$ can be taken as the plate separation. In this case, a *true characteristic field enhancement factor (FEF)* $\gamma_C$ is defined and given adequately by

$$\gamma_C \equiv F_C / F_M = \zeta_M/\zeta_C. \quad (63)$$

### D. Area-like quantities

There are also auxiliary equations, related to eq. (60), that define area-like quantities. A formula for *emission current* $i_e$ can be obtained formally by integrating the local current density $J_L$ over the whole emitting surface and writing the result in the form:

$$i_e = \int J_L dA = A_n J_C. \quad (64)$$

The parameter $A_n$ is termed the *notional emission area* and relates to the area of the emitter that is actually emitting.

Using eq. (58) (but dropping the label "GB"), eq. (64) can also be written in the equivalent forms

$$i_e = A_n J_C = A_n \lambda_C J_{kC} \equiv A_f J_{kC}, \quad (65)$$

where $A_f [\equiv A_n \lambda_C]$ is the *formal emission area*. It is this area parameter that is extracted from an appropriate FN plot (if the emission situation is orthodox) [22].

For a large area field electron emitter (LAFE), the *macroscopic (or "average") current density* $J_M$ is given by

$$J_M = i_e/A_M = (A_n/A_M)J_C = \alpha_n J_C, \quad (66)$$

where $A_M$ is the *macroscopic area (or "total footprint")* of the LAFE, and $\alpha_n$ is called its *notional area efficiency*. As before, we also have the corresponding formal parameter, given via

$$J_M = \alpha_n J_C = \alpha_n \lambda_C J_{kC} \equiv \alpha_f J_{kC}, \quad (67)$$

where $\alpha_f [\equiv \alpha_n \lambda_C]$ is the *formal emission area*.

At present, actual values of these area-like quantities are not well known. In principle, values of $A_f$ and $\alpha_f$ could be extracted from orthodox FN plots, but this is not done in most experiments.

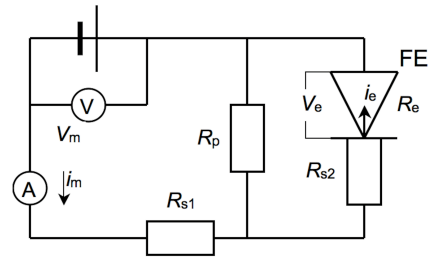

FIG. 1. Schematic electric circuit for measuring the current-voltage characteristics of a field electron emitter.

### E. Emission variables and measured variables

A schematic FE measurement circuit is shown in Fig. 1. In this circuit, the *emission voltage* $V_e$ is the voltage between the emitting regions at the emitter tip and the counter-electrode, the *emission current* $i_e$ is the current though the emitting device, and the *emission resistance* $R_e$ is defined by

$$R_e = V_e/i_e. \quad (68)$$

In real measurement circuits, $V_e$ and $i_e$ may not be equal to the *measured voltage* $V_m$ and the *measured current* $i_m$

(measured in the vicinity of the high-voltage generator), due to the presence of series and/or parallel resistance. The parallel resistance $R_p$ can usually be made sufficiently large to ensure no significant leakage current, but the resulting series resistance $R_s$ [$=R_{s1}+R_{s2}$] often cannot be eliminated.

For a field emitter, the emission resistance $R_e$ is strongly current dependent, being very large at low emission voltages, and getting smaller as voltage increases. When $R_e$ begins to become comparable with $R_s$, "voltage loss" effects will set in, and the emission voltage will no longer equal the measured voltage. As a result, the measured current and voltage will no longer have the relationship expected from a FN-type equation (which, of course, applies physically to the emission variables).

*F. Fowler-Nordheim plots and the Orthodoxy Test*

As is well known, a plot of the form $\ln\{i_m/V_m\}$ vs $1/V_m$ is called a *Fowler-Nordheim plot (FN plot)*, and (for metal emitters) is expected to be a straight line. In favorable circumstances, emitter characterization parameters can be deduced from the slope and intercept of the FN plot. The details of how to do this have recently been reviewed [22], and will not be repeated here.

The conditions under which analysis of this kind is expected to be valid have recently been formulated as a set of *orthodoxy conditions* [21], and an *orthodoxy test* has been invented [21] that can test whether or not a particular FN plot relates to an orthodox emission situation.

Preliminary checks suggest that FN plots in a significant percentage of published papers may fail the orthodoxy test, and consequently that many of these papers may be reporting spuriously high field-enhancement-factor (FEF) values. These issues, and some possible methods of dealing with them, have been recently been discussed in detail elsewhere [21,22].

*G. Some issues not covered*

In the time available in the tutorials, it was not possible to cover all relevant issues. Important mainstream issues not covered include: energy distribution theory; Henderson/Nottingham emitter cooling and heating effects; details of interpreting current-voltage data for LAFEs; electrostatic interactions between emitters; and theory related to the fabrication, degradation, and regeneration of emitters.

There is also some very interesting work going on outside what I have described as mainstream theory — notably work on detailed theories of emission from carbon nanotubes, and work on laser-pulsed field electron emission and related effects. There is also significant activity in understanding the role of FE in vacuum breakdown, particularly in the context of high-voltage-gradient accelerators.

VII. SOME OUTSTANDING PROBLEMS AND TASKS

A principal aim of the author's research activity is to improve the basic science of field electron emission and its clear communication amongst scientists. The second tutorial [2] ended with an assessment of outstanding problems in and close to mainstream FE science, and a suggested list of 16 "immediately outstanding" tasks. These are as follows.

1. Encourage all FE work to be presented using only the International System of Quantities [i.e., abandon 1960s style Gaussian-system equations and related "hybrid" conventions].

2. Encourage the LAFE community to abandon use of the elementary FN-type equation in data analysis, in favor of a system based on the orthodox FN-type equation.

3. Encourage standardization of terminology and notation.

4. Encourage use of the Gauss variable $x$, rather than the Nordheim parameter $y$, in the theory of the SN barrier.

5. Develop a single coherent approach to extracting formal emission area from orthodox FN plots.

6. Develop further the theory of data analysis in non-orthodox emission situations.

7. By using the orthodoxy test, investigate the extent of the "spurious results" problem in FE literature, and also whether useful information can be extracted by re-analyzing the data in these and other published "emitter characterization" papers.

8. Find means of investigating empirically whether the classical image PE is a satisfactory approximate model for the exchange-and-correlation interaction between an escaping electron and the emitter.

9. Investigate experimentally what the actual power of voltage is in FN-type equation pre-exponentials.

10. Find means of making experimental estimates of the value of the pre-exponential correction factor $\lambda_C$.

11. Investigate further the theory of transmission near the top of the SN barrier.

12. Integrate better into mainstream theory Jensen's more-general "temperature-field" formula [14,15].

13. Establish improved methods of defining emission regimes, and re-investigate Murphy-Good theory for the boundaries of the FNFE (CFE) regime.

14. Investigate the validity of JWKB-type methods of evaluating transmission probability when the Schrödinger equation does not separate in Cartesian coordinates.

15. Given that the field electron microscope can "see carbon bonds", investigate further the theory of FEM resolution.

16. Attempt to relate the theory of FE from carbon nanotubes more closely with mainstream FE theory.

NOTE ADDED IN MAY 2017

The original version of this paper has now been published on-line [24]. The present version is a slightly updated version, in which the main change is a more careful discussion of the parameter $\tau_F^{SN}$. Note also that it is now thought [25] that, with LAFEs, the principal cause of non-orthodoxy effects in FN plots may be current-dependence in field enhancement factors, rather than series resistance in the measurement circuit.

The two tutorials to which this paper originally related have recently (in April 2017) been presented in an updated and extended form, as part of a recent conference, and the related power-point presentations have been uploaded to ResearchGate [26, 27]. Those interested are referred to these later presentations.